# Effects of Sequence Partitioning on Compression Rate

B. Baykant ALAGOZ

**Abstract:** *In the paper, a theoretical work is done for investigating effects of splitting data sequence into packs of data set. We proved that a partitioning of data sequence is possible to find such that the entropy rate at each subsequence is lower than entropy rate of the source. Effects of sequence partitioning on overall compression rate are argued on the bases of partitioning statistics, and then, an optimization problem for an optimal partition is defined to improve overall compression rate of a sequence.*

## 1. Introduction:

Shannon established that there is a fundamental limit to lossless data compression in 1948. This limit is called as entropy rate. The exact value of entropy rate depends on the statistical nature of the source. Shannon stated that it was possible to compress information from a data source, in a lossless manner, with compression rate close to the entropy rate by using appropriate coding method.[1-2] The most popular coding methods are Huffman Coding and Lempel-Ziv Coding. [3-5] Depending on the statistical order of data source, entropy rate of the source decreases, and thus, a better compression rate can be reachable by developing appropriate coding methods.

The partitioning of a data sequence was practically shown to provide a better compression rate.[6] In Huffman coding scheme, splitting a original symbol sequence into sub-sequences was shown to give a better bit rates for compression of AR1, ECG and seismic signals at several SNR. The developed recursive splitting method partitions an original symbol sequence in a way that it can make the symbol probabilities different for each subsequence, where Huffman coding exhibits a better performance. In this fashion, the following important question comes to minds: does a sequence partitioning works for the improvement of compression performance of other coding methods? How does a partition affect the compression rate? What is the lower bound for overall compression rate of a partitioned sequence? We will try to find answers of these questions in this paper.

Preliminarily, we aimed to demonstrate that there can be always found a partition that makes entropy rate at each pack of data lower than original data sequence. In proceeding, we give a brief analysis on overall compression rate of a partitioned data sequence and inspect factors affecting this rate. A lower bound for the overall compression rate of a partitioned data sequence is given in the light of Shannon's limit.

Uncovering dynamics behind sequence partitioning in the term of data compression will help the practical efforts to improve compression performance of conventional coding schemes. An optimization problem basing on those dynamics are defined for development of practical application.

## 2. Method
### 2.1 Decreasing Overall Entropy Rates of a Infinite Data Stream by Partition into Lower Entropy Pack of Subsets:

In this section, a data sequence is theoretically shown to be partitioned into subsequences such that entropy rate of each sequence is smaller than the entropy rate of original sequence.

*Theorem 1:* Let an infinite set of data be $X = \{x_1, x_2....x_l...\}$ and it is generated by a zero-order source. Let a finite set of symbol be $A = \{a_1, a_2....a_m\}$, where $m > 1$, and the data set $X$ is composed of elements of symbol set $A$. A partition of data set $X$ can always be found such that entropy rate in each subset is lower than entropy of rate in the set $X$.

*Proof:*
One always can form a subset of $X$ from the first $m-1$ elements of $X$. Lets denote this subset by $X_1$. For a zero-order source model, entropy rate $H$ for $X$ can be written as, [1]

$$H = \log_2 m. \qquad (1)$$

For the subset $X_1$; since it has $m-1$ elements, symbol set of $X_1$ never becomes larger than $m-1$. It specifies the upper bound for entropy rate at $X_1$. In this case, entropy rate for $X_1$ is written as

$$H_1 \leq \log_2(m-1) \qquad (2)$$

Therefore $H_1 < H$, one can clearly state that at least a subset of the data set $X$ can always be found such that entropy rate at subset is lower than entropy rate at the set $X$.

Let consider next $m-1$ elements of $X$ form the subset $X_2$ and the following $m-1$ elements of $X$ forms the subset $X_3$ and so on. Now, one have $X_1, X_2, X_3,...$ subset family of $X$, such that the entropy rate at each family member $X_i$ is lower than the entropy rate at $X$. Since $X = X_1 \cup X_2 \cup X_3 \cup ......$ and all $H_i < H$, one can state that a partition of the data set $X$ can always be found such that the entropy rate of each subset is lower than entropy rate of the set $X$.

*Theorem 2:* When an finite set of data $X = \{x_1, x_2....x_l\}$ from a zero-order model source is split into subsets $X_1, X_2,..., X_g$, where $g \in [2, l]$. Entropy rate at each subset is equal or lower than entropy rate at $X$.

*Proof:*
If $X$ is a finite set, the symbol set of $X$, $A$, has to be finite as well. Let the number of elements in $A$ denote by $m$. The entropy rate at $X$ can be written as $H = \log_2 m$. Since $A_i$ symbol set of a subset $X_i$ is contained in $A$, number of element in any subset $A_i$ will be equal or lower than $m$. Therefore, entropy rate at a subset $X_i$, denoted by $H_i$, will be equal and lower than $H$. So, $H_i \leq H$.

Theorem 1 tell us that the partitioning works well for reducing entropy rates in data stream, however, it does not imply that a better overall compression rate can be reached by using such a partitioning. In the following section, we will derive an expression for the overall compression rate of partitioned sequences.

## 2.2 Overall Compression Rate of a Partitioned Data Sequence

In order to make an analysis of how the partitioning of a finite data sequence affects the overall compression rate, firstly let figure out the overall compression rate of a finite data sequence $D$, in the case that it is split into $k$ number of subsequence denoted by $d_i$, $i \in [1,k]$. Each subsequent $d_i$ is assumed to have the number of element denoted by $a_i$. Overall compression rate of $D$ in this partition can be written as,

$$R = \sum_{i=1}^{k} w(a_i) \cdot R_i, \qquad (3)$$

where $R_i$ is the compression rate obtained in coding subsequent $d_i$ by mean of any coding schemes and $w(a_i)$ is size weight of $d_i$ in $D$ set and expressed as $w(a_i) = a_i / l$. Here, $l$ is number of element in $D$. (See the appendix for derivation of equation (3).)

Since $R_i > H_i$ according to Shannon's limits of lossless compression,[1-2] an lower bound for the overall compression rate appears as,

$$R > \sum_{i=1}^{k} w(a_i) \cdot H_i. \qquad (4)$$

Considering equations (3) and (4), we can list the following remarks:
- Overall compression rate of a partitioned sequence depends on compression rate of each subsequence and their sizes. In order to reach a lower overall compression, one should establish an optimal partition strategy such that the subsequences exhibiting a lower compression rates have the larger in size.
- It is not a necessity to use one type coding schemes for coding all subsequences. In the multiple-coding approach, the one providing a lowest rate of compression among the coding schemes is selected to code for this subsequence of partitioning.
- Considering the condition of $H_i \leq H$ from Theorem 2 and the equation (4), we reaches such a important conclusion that the sequence partitioning is preferable for improving overall compression rate of original sequence, because of its potential to reduce the lower bound of overall compression rate.

## 3 A Discussion for Improving Overall Compression Rate by Optimal Partitioning of Data Sequence

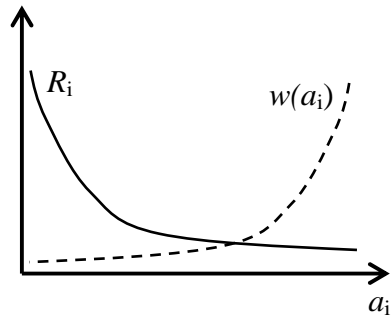

**Figure 1.** Distribution of $R_i$ and $w(a_i)$ that is improving overall compression rate in partitioning of a sequence.

In order to obtain a lower overall compression ratio, a partition characteristic illustrated in Figure 1 has to be aimed in splitting original sequence. According to this characteristic, following partitioning rules, we call it as entropic partitioning, can be derived:

- For the data segments having higher entropy rates, form subsequences with shorter length,
- For the data segments having lower entropy rates, form subsequences with larger length as in Figure 2.

Entropic partitioning is not dependent of coding methods. A good partition, for the case of a predetermined set of coding schemes, is argued below:

In Figure 3, an example of good partition is illustrated on a time series signal. An optimal partitioning of data sequence can be defined as a partitioning that makes the overall compression rate a globally minimum, and it is simply expressed as

$$R_{opt} = \min_{a_i, R_i} (\sum_{i=1}^{k} w(a_i) \cdot R_i). \tag{5}$$

For the practice point of view, the problem of finding an optimal partitioning of a sequence turns into problem of finding $a_i, R_i$ parameters that yields a minimal overall compression rate. An objective function to be optimized may be fashionably given as,

$$E = (\sum_{i=1}^{k} w(a_i) \cdot R_i)^2. \tag{6}$$

For a quick approximation to the optimal solution characteristics presented by Figure 1, a partitioning can be performed subject to a constant bit-length constraint, which is arithmetically defined as,

$$C = a_i \cdot R_i, \tag{7}$$

where $C \in R$ is a constant implying a target length for subsequences in term of bits. An advanced version of the objective function to be optimized for an optimal partitioning with a constant bit-length of subsequences can be given as,

$$E = \sum_{i=1}^{k} [(w(a_i) \cdot R_i)^2 + (C - a_i \cdot R_i)^2]. \tag{8}$$

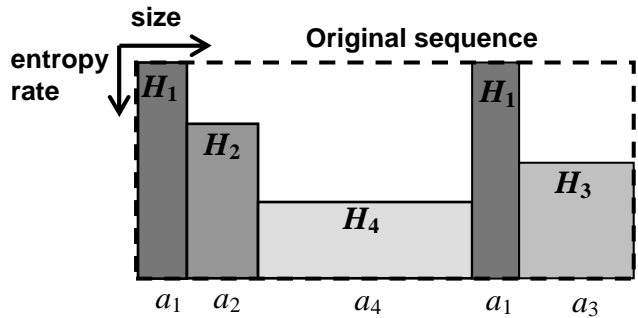

**Figure 2.** An example of good partitioning: Original sequence, represented by a large dash rectangle, have an entropy rate $H_1$ and the size of $2a_1 + a_2 + a_3 + a_4$. Subsequences are represented by rectangular areas in gray.

Introduction of an algorithm to solve in this optimization problem is not a preference of this work. We rather aim to define an optimization problem for the optimal partitioning, which is applicable in all practical coding schemes.

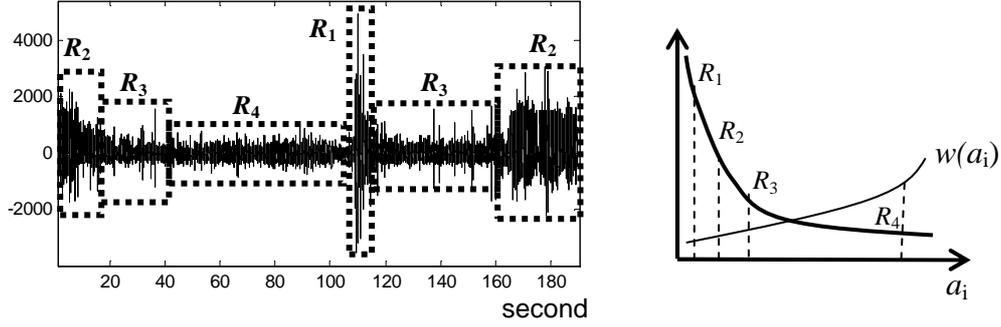

**Figure 3.** An example illustrating a good partitioning on a time series signal.

### 4. Conclusions:

The entropy rate at a data sequence can be easily decreased by splitting it. However, the overall compression rate of a partitioned sequence is seen to be dependent on compression rate of coding methods applied to subsequences and size of subsequences. In the paper, a theoretical discussion for a general optimal partition strategy, which is applicable to all coding techniques, was given and an optimization problem is defined to improve compression rate of data sequence. Furthermore, we see that it may be possible to use a collaboration of coding method in a partitioned sequence to reach a better compression performance. We referred it as multi-coding optimal partitioning.

With findings of this theoretical work, roles of partitioning in coding will be better understood and hopefully practical attempts to develop novel methods for the optimal partitioning of a data sequence can be enthused in the field.

### Appendix:

*Derivation of overall compression rate formula*:

A finite sequence with $l$ elements splits into $k$ number of subsequences, which have the lengths of $a_i$ and the compression rate of $R_i$. In this case, the total bit number used for coding this sequence can be written as,

$$T = \sum_{i=1}^{k} a_i \cdot R_i .$$

Compression rate, defined as number of bit per symbol in sequence, can be expressed as $R = T/l$. So, the overall compression rate for a partitioned sequence is written as,

$$R = \sum_{i=1}^{k} \frac{a_i}{k} \cdot R_i .$$

$w(a_i) = a_i / l$, is used, one obtains overall compression rate as:

$$R = \sum_{i=1}^{k} w(a_i) \cdot R_i .$$